\documentstyle[12pt]{article}
\newfont{\bbd}{msbm10 scaled\magstep1}

\begin{document}
\thispagestyle{empty}

\newcommand{\p}[1]{(\ref{#1})}
\newcommand{\be}{\begin{equation}}
\newcommand{\ee}{\end{equation}}
\newcommand{\sect}[1]{\setcounter{equation}{0}\section{#1}}

\newcommand{\vs}[1]{\rule[- #1 mm]{0mm}{#1 mm}}
\newcommand{\hs}[1]{\hspace{#1mm}}
\newcommand{\mb}[1]{\hs{5}\mbox{#1}\hs{5}}
\newcommand{\Db}{{\overline D}}
\newcommand{\bea}{\begin{eqnarray}}
\newcommand{\eea}{\end{eqnarray}}
\newcommand{\wt}[1]{\widetilde{#1}}
\newcommand{\und}[1]{\underline{#1}}
\newcommand{\ov}[1]{\overline{#1}}
\newcommand{\sm}[2]{\frac{\mbox{\footnotesize #1}\vs{-2}}
		   {\vs{-2}\mbox{\footnotesize #2}}}
\newcommand{\prt}{\partial}
\newcommand{\eps}{\epsilon}

\newcommand{\R}{\mbox{\rule{0.2mm}{2.8mm}\hspace{-1.5mm} R}}
\newcommand{\Z}{Z\hspace{-2mm}Z}

\newcommand{\cd}{{\cal D}}
\newcommand{\cg}{{\cal G}}
\newcommand{\ck}{{\cal K}}
\newcommand{\cw}{{\cal W}}

\newcommand{\vj}{\vec{J}}
\newcommand{\vl}{\vec{\lambda}}
\newcommand{\vz}{\vec{\sigma}}
\newcommand{\vt}{\vec{\tau}}
\newcommand{\vw}{\vec{W}}
\newcommand{\poiss}{\stackrel{\otimes}{,}}

\def\l#1#2{\raisebox{.2ex}{$\displaystyle
  \mathop{#1}^{{\scriptstyle #2}\rightarrow}$}}
\def\r#1#2{\raisebox{.2ex}{$\displaystyle
 \mathop{#1}^{\leftarrow {\scriptstyle #2}}$}}

\newcommand{\NP}[1]{Nucl.\ Phys.\ {\bf #1}}
\newcommand{\PL}[1]{Phys.\ Lett.\ {\bf #1}}
\newcommand{\NC}[1]{Nuovo Cimento {\bf #1}}
\newcommand{\CMP}[1]{Comm.\ Math.\ Phys.\ {\bf #1}}
\newcommand{\PR}[1]{Phys.\ Rev.\ {\bf #1}}
\newcommand{\PRL}[1]{Phys.\ Rev.\ Lett.\ {\bf #1}}
\newcommand{\MPL}[1]{Mod.\ Phys.\ Lett.\ {\bf #1}}
\newcommand{\BLMS}[1]{Bull.\ London Math.\ Soc.\ {\bf #1}}
\newcommand{\IJMP}[1]{Int.\ Jour.\ of\ Mod.\ Phys.\ {\bf #1}}
\newcommand{\JMP}[1]{Jour.\ of\ Math.\ Phys.\ {\bf #1}}
\newcommand{\LMP}[1]{Lett.\ in\ Math.\ Phys.\ {\bf #1}}

\renewcommand{\thefootnote}{\fnsymbol{footnote}}
\newpage
\setcounter{page}{0}
\pagestyle{empty}
\begin{flushright}
{January 2002}\\
{nlin.SI/0201026}\\
\end{flushright}
\vs{8}
\begin{center}
{\LARGE {\bf The N=2 supersymmetric unconstrained}}\\[0.6cm]
{\LARGE {\bf matrix GNLS hierarchies}}\\[1cm]

\vs{8}

{\large A.S. Sorin$^{(a)}$ and P.H.M. Kersten$^{(b)}$}
{}~\\
\quad \\
{\em {~$~^{(b)}$ Bogoliubov Laboratory of Theoretical Physics, \\
Joint Institute for Nuclear Research (JINR),}}\\
{\em 141980 Dubna, Moscow Region, Russia \\
E-Mail: sorin@thsun1.jinr.ru}\\
{}~\\
{\em ~$~^{(a)}$ University of Twente, Faculty of Mathematical Sciences, \\
P.O.Box 217,7500 AE Enschede, The Netherlands \\
E-Mail: kersten@math.utwente.nl}~\quad\\

\end{center}
\vs{8}

 \centerline{ {\bf Abstract}}
The generalization of the $N=2$ supersymmetric chiral matrix 
$(k|n,m)$--GNLS hierarchy (Lett. Math. Phys. 45 (1998) 63,
solv-int/9711009) to the case when matrix entries are bosonic and
fermionic unconstrained $N=2$ superfields is proposed. 
This is done by exhibiting the corresponding matrix Lax--pair
representation in terms of $N=2$ unconstrained superfields. It is 
demonstrated that when matrix entries
are chiral and antichiral $N=2$ superfields, it reproduces the $N=2$
chiral matrix $(k|n,m)$-GNLS hierarchy, while in the 
scalar case, $k=1$, it is equivalent to the $N=2$ supersymmetric
multicomponent hierarchy (J. Phys. A29 (1996) 1281, hep-th/9510185). 
The simplest example --- the $N=2$ unconstrained  $(1|1,0)$--GNLS
hierarchy --- and its reduction to the $N=2$ supersymmetric
${\alpha}=1$ KdV hierarchy are discussed in more detail, and its rich
symmetry structure is uncovered. 

{}~

{}~

{}~



\newpage

\pagestyle{plain}
\renewcommand{\thefootnote}{\arabic{footnote}}
\setcounter{footnote}{0}

\section{\bf Introduction}
The $N=2$ chiral matrix $(k|n,m)$--Generalized Nonlinear 
Schr\"{o}dinger (MGNLS) hierarchies were introduced in \cite{bks1}.
Their group--theoretical origin was clarified in \cite{bks2} where 
it has been demonstrated that they are related to the $N=2$ $sl(n|n-1)$
affine superalgebras via the coset construction. The $N=2$ chiral
$(k|n,m)$--MGNLS hierarchies comprise rectangular matrix-valued 
{\it constrained} (chiral and antichiral) bosonic and fermionic $N=2$
superfields. The aim of the present Letter is to construct 
their {\it integrable} generalizations in the case when matrix
entries are {\it unconstrained} $N=2$ superfields. It turns out that such
generalizations indeed exist, and we call them the $N=2$ supersymmetric
unconstrained $(k|n,m)$--MGNLS hierarchies in what follows.

The Letter is organized as follows. In Section 2 we introduce
the $N=2$ unconstrained $(k|n,m)$--MGNLS hierarchy, discuss its 
properties and construct relevant quantities. In Section 3 we analyse its
bosonic limit as well as a correspondence of some of its limited cases to
known hierarchies. In particular, we discuss the reduction of the $N=2$
unconstrained $(1|1,0)$--MGNLS hierarchy to the $N=2$ supersymmetric
${\alpha}=1$ KdV hierarchy \cite{lm,pop1} and uncover a rich symmetry
structure of the latter. In Section 4 we summarize our results.

\section{\bf The $N=2$ unconstrained $(k|n,m)$--MGNLS hierarchies}

In this section we introduce the Lax--pair representation of the 
$N=2$ unconstrained $(k|n,m)$--MGNLS hierarchy, construct its
conserved quantities and different complex conjugations. 
  
\subsection{\bf Lax--pair representation}

We propose the following Lax--pair representation for the bosonic flows
of the $N=2$ supersymmetric unconstrained $(k|n,m)$--MGNLS hierarchies:  
\begin{eqnarray}
{\textstyle{\partial\over\partial t_p}}L =[ A_p , L],\quad
L = I\partial + \frac{1}{2}F D{\overline D}\partial^{-1} {\overline F},
\quad  A_p = (L^p)_{\geq 0} + res(L^p), \quad p \in \hbox{\bbd N}
\label{suplax}
\end{eqnarray}
generating the abelian algebra of the flows
\begin{eqnarray}
[{\textstyle{\partial\over\partial t_m}},
{\textstyle{\partial\over\partial t_n}}]=0
\label{alg}
\end{eqnarray}
where the subscript $\geq 0$ denotes the 
sum of purely differential and constant parts of the operator
$L^p$, and $res(L^p)$ is its $N=2$ supersymmetric residue, i.e.
the coefficient at $[D,\overline D]{\partial}^{-1}$.
Here, $F\equiv F_{Aa}(Z)$ and ${\overline F}\equiv {\overline F}_{aA}(Z)$
($A,B=1,\ldots, k$; $a,b=1,\ldots , n+m$) are rectangular 
matrices which entries are unconstrained $N=2$ superfields, $I$ is the
unity matrix, $I\equiv {\delta}_{A,B}$, and the matrix product is
understood, for example, as $(F\overline F)_{AB} \equiv \sum_a
F_{Aa}\overline F_{aB}$. The matrix entries are bosonic superfields for
$a=1,\ldots ,n$ and fermionic superfields for $a=n+1,\ldots , n+m$, 
i.e., $F_{Aa}{\overline F}_{bB}=(-1)^{d_{a}{\overline d}_{b}}
{\overline F}_{bB}F_{Aa}$ where $d_{a}$ and ${\overline d}_{b}$ are the
Grassmann parities of the matrix elements $F_{Aa}$ and ${\overline F}_{bB}$,
respectively, $d_{a}=1$ $(d_{a}=0)$ for fermionic (bosonic) entries;
$Z=(z,\theta,\overline\theta)$ is a coordinate in the $N=2$ superspace,
$dZ \equiv dz d \theta d \overline\theta$ and $D,{\overline D}$ are the
$N=2$ supersymmetric fermionic covariant derivatives
\begin{eqnarray}
D=\frac{\partial}{\partial\theta}
 -\frac{1}{2}\overline\theta\frac{\partial}{\partial z}, \quad
{\overline D}=\frac{\partial}{\partial\overline\theta}
 -\frac{1}{2}\theta\frac{\partial}{\partial z}, \quad
D^{2}={\overline D}^{2}=0, \quad
\left\{ D,{\overline D} \right\}= -\frac{\partial}{\partial z}
\equiv -{\partial}.
\label{DD}
\end{eqnarray}
The chosen grading guarantees that the Lax operator $L$ \p{suplax} is
Grassman even \cite{bks1}. 

We have verified for first few non-trivial flows that the
Lax--pair representation \p{suplax} consistently provides their locality
and leads to the following general flow equations for $F$ and
$\overline F$:
\begin{eqnarray}
{\textstyle{\partial\over\partial t_p}}F =
(A_{p} F)_0, \quad 
{\textstyle{\partial\over\partial t_p}} {\overline F}^{T} =
-(A_{p}^{T}{\overline F}^{T})_0
\label{n1}
\end{eqnarray}
which explicitly are\footnote{Hereafter, the subscript
$0$ denotes the constant part of the corresponding operators, and 
we also use the notation $(\hbox{\bbd O}f)$ for an operator 
$\hbox{\bbd O}$ acting only on a function $f$ inside the brackets.}: 
\begin{eqnarray} 
&&{\textstyle{\partial\over\partial t_0}} F = F,\quad
{\textstyle{\partial\over \partial t_0}} {\overline F} =
-{\overline F}; \quad
{\textstyle{\partial\over\partial t_1}} F =F~', \quad
{\textstyle{\partial\over \partial t_1}} {\overline F} =
{\overline F}~'; \nonumber\\
&&{\textstyle{\partial\over\partial t_2}} F =
F~'' +  F (D{\overline D}~{\overline F} F), \quad
{\textstyle{\partial\over \partial t_2}} {\overline F} =
-{\overline F}~'' + ({\overline D}{D}{\overline F}F) 
{\overline F};\nonumber\\
&&{\textstyle{\partial\over\partial t_3}} F =
F~''' + \frac{3}{2}[ F~' (D{\overline D}~{\overline F} F)+
F (D{\overline D}~{\overline F} F~') - \frac{1}{2}
F (D{\cal I}^2{\overline F} F {\overline D}~{\cal I}^2{\overline F} F)], 
\nonumber\\ &&{\textstyle{\partial\over \partial t_3}} {\overline F} =
{\overline F}~''' -\frac{3}{2} 
[({\overline D}{D}~{\overline F}F){\overline F}~'+
({\overline D}{D}~{\overline F}~'F){\overline F}+\frac{1}{2}
({\overline D}({D}{\overline F}F){\overline F}F){\overline F}].
\label{gnls}
\end{eqnarray}
Here, we have introduced the matrix ${\cal I}$, 
\begin{eqnarray}
{\cal I} \equiv (i)^{{\overline d}_a}{\delta}_{ab}, \quad 
{\cal I}^2 = (-1)^{{\overline d}_a}{\delta}_{ab}, \quad 
{\cal I}^4=I,
\label{matrixI}
\end{eqnarray}
$'$ denotes the derivative with respect to $z$ and
we use the following standard convention regarding the 
operator conjugation (transposition) $^T$
\begin{eqnarray}
&&({\partial},D,{\overline D})^{T}=-({\partial},D,{\overline D}),\quad
(D{\overline D})^{T} = - {\overline D}D, \nonumber\\ 
&& (\sum_{a} F_{Aa}{\overline F}_{aB})^{T}=\sum_{a}(-1)^{d_{a}}
({\overline F}_{aB})^{T} (F_{Aa})^{T}. 
\label{transp}
\end{eqnarray}
For completeness, let us also present expressions for the
corresponding operators $A_p$, $res(L^p)$ and $(L^p)_{0}$,  
\begin{eqnarray}
&&A_0= 1, \quad A_1= {\partial}, \quad
A_2=I{\partial}^2+FD{\overline D}~{\overline F}, \nonumber\\
&&  ~A_3=I{\partial}^3+\frac{3}{2}F~'D{\overline D}~{\overline F}+
\frac{3}{2}FD{\overline D}~{\overline F}{\partial}-
\frac{3}{4}FD{\cal I}^2{\overline F}F{\overline D}{\cal I}^2{\overline F}
\label{laxfl1}
\end{eqnarray}
and
\begin{eqnarray}
&&res(L)= \frac{1}{4} F{\overline F}, \quad
 res(L^2)= \frac{1}{4} (F~'{\overline F}-
F{\overline F}~'-\frac{1}{2}(F{\overline F})^2), \nonumber\\ 
&&res(L^3) = \frac{1}{4}(F{\overline F})~''+\frac{1}{8}
[F{\overline F},(F{\overline F})~']
\nonumber\\&&  ~~~~~~~~~~~+\frac{3}{4}(-F~'{\overline F}~'+
\frac{1}{2} F(D{\overline D}~ {\overline F}F)_0 {\overline F} +\frac{1}{2}
F {\overline F}~' F {\overline F} + \frac{1}{12} (F {\overline F})^3)
\label{rese}
\end{eqnarray}
as well as
\begin{eqnarray}
&&L_0= -res(L), \quad
(L^2)_{0}= F(D{\overline D}~{\overline F})_{0}- res(L^2),
\nonumber\\ &&(L^3)_{0} ~~= \frac{3}{2}F~'
(D{\overline D}~{\overline F})_{0} -
\frac{3}{4}F(D{\cal I}^2{\overline F}F{\overline D}{\cal I}^2
{\overline F})_{0} - res(L^3),
\label{help}
\end{eqnarray}
respectively, which will be useful in what follows. 

Let us note that besides the global $N=2$ supersymmetry the Lax operator
$L$ \p{suplax} and the flows (\ref{n1}--\ref{gnls}) of the $N=2$
unconstrained $(k|n,m)$--MGNLS hierarchy are obviously invariant with
respect to the direct product of the (super)groups $GL(k)\times GL(n|m)$,
and the matrices $L_{AB}$, (${\overline F}_{aB}$) $F_{Ab}$ realize their
(anti)fundamental representations over the $GL(k)$--indices 
($B$) $A$ and the $GL(n|m)$--indices ($b$) $a$. 

The Lax operator $L$ \p{suplax} contains the constant part over the
derivative ${\partial}$
\begin{eqnarray}
L_{0} =-\frac{1}{4}F {\overline F}. 
\label{const}
\end{eqnarray}
Let us discuss another, gauge--related Lax--pair representation 
\begin{eqnarray}
{\textstyle{\partial\over\partial t_p}} {\widetilde L} = 
[ {\widetilde A}_p , {\widetilde L}],\quad
{\widetilde L} = G LG^{-1}
\quad  {\widetilde A}_p = 
G\Bigl(A_p + G^{-1}{\textstyle{\partial\over\partial t_p}}G\Bigr)G^{-1} 
\label{gauge}
\end{eqnarray}
which is fixed by a requirement that the gauge--transformed Lax operator 
${\widetilde L}$ does not contain the constant part. The latter 
leads to the following equation for the matrix 
of the gauge transformation $G\equiv (G)_{AB}$: 
\begin{eqnarray}
G^{-1}G~' = L_0. 
\label{Ggauge}
\end{eqnarray}
In order to find an equation for the quantity  
$G^{-1}{\textstyle{\partial\over\partial t_p}}G$ entering into 
eq. \p{gauge} we differentiate eq. \p{Ggauge} over 
${\textstyle{\partial\over\partial t_p}}$, then substitute 
\begin{eqnarray}
{\textstyle{\partial\over\partial t_p}}L_0 =
-(res(L^p))~' +2[res(L^p), L_0 ]
\label{suplaxconst}
\end{eqnarray}
which results from the constant part of the Lax--pair
representation \p{suplax}, and finally have
\begin{eqnarray}
(G^{-1}{\textstyle{\partial\over\partial t_p}}G)~' 
-[G^{-1}{\textstyle{\partial\over\partial t_p}}G, L_0 ]=
-(res(L^p))~'  + 2[res(L^p), L_0 ].
\label{eqG}
\end{eqnarray}
A simple inspection of this equation shows that it has an obvious
local solution
\begin{eqnarray}
G^{-1}{\textstyle{\partial\over\partial t_p}}G= -res(L^p) 
\label{solG}
\end{eqnarray}
in the commutative, scalar case, i.e. for the $N=2$ supersymmetric
unconstrained $(1|n,m)$--MGNLS hierarchies, and, consequently, the
operator ${\widetilde A}_p$ \p{gauge} becomes
\begin{eqnarray}
{\widetilde A}_p = ({\widetilde L}^p)_{\geq 0}
\label{Aop1}
\end{eqnarray}
using the identity
\begin{eqnarray}
G (L^p)_{\geq 0}G^{-1}\equiv (G L^pG^{-1})_{\geq 0}.
\label{Aop2}
\end{eqnarray}
Though for the noncommutative, matrix case, i.e. for the $N=2$
unconstrained $(k \geq 2|n,m)$--MGNLS hierarchies, the solution 
for $G^{-1}{\textstyle{\partial\over\partial t_p}}G$ is
nonlocal in general, and as a consequence the Lax--pair representation
\p{gauge} is nonlocal as well. In this respect the {\it local} Lax--pair
representation \p{suplax} we started with is rather {\it exceptional}.

To close this subsection, we would like to remark that integrability
conditions \p{alg} for the equations \p{suplaxconst} read
\begin{eqnarray}
{\textstyle{\partial\over\partial t_n}} res(L^m)-
{\textstyle{\partial\over\partial t_m}} res(L^n) + 2[res(L^m),res(L^n)]=0,
\label{integrcond}
\end{eqnarray}
therefore, $res(L^m)$ can consistently be represented in terms of the
single matrix $X\equiv X_{AB}$ 
\begin{eqnarray}
res(L^m)=-\frac{1}{2}X^{-1}{\textstyle{\partial\over\partial t_m}} X 
\label{def}
\end{eqnarray}
which will be useful in what follows (see, subsection 2.3).

\subsection{\bf Hamiltonians}
The infinite set of Hamiltonians can be defined as:
\begin{eqnarray}
{H}_p =4\int d Z ~{\cal H}_{p}, \quad 
{\cal H}_{p}\equiv tr(res(L^p))
\label{res}
\end{eqnarray}
where $tr$ is the usual matrix trace. Their conservation is the
obvious consequence of the Lax--pair representation \p{suplax}.
By construction, these Hamiltonians
presumably correspond to the flows 
${\textstyle{\partial\over\partial t_p}}$
\p{suplax} via the corresponding Hamiltonian structure (if any) as
usually, and in this
case they have to form an abelian algebra because of the well--known
homomorphism between algebra of flows, which is the abelian algebra
\p{alg} in the case under consideration, and algebra of their 
Hamiltonians.
Substituting expressions for $res(L^p)$ \p{rese} into eq. \p{res} one can
obtain few first Hamiltonians from the set \p{res}  
\begin{eqnarray}
&&\quad \quad
H_1 = \int dZ ~tr (F {\overline F}), \quad
H_2 = -2\int dZ ~tr ( F {\overline F}~' 
+ \frac{1}{4} (F{\overline F})^2), \nonumber\\&& 
H_3 = 3\int dZ~tr ( F {\overline F}~'' +
\frac{1}{2} F(D{\overline D}~ {\overline F}F) {\overline F} +\frac{1}{2}
F {\overline F}~' F {\overline F} + \frac{1}{12} (F {\overline F})^3).
\label{i1}
\end{eqnarray}

Besides the Hamiltonians $H_p$ \p{res} there exist other conserved
quantities of the flows \p{gnls} which presumably form a non-abelian
algebra. Thus, one can verify by direct calculation that the supermatrix
functionals
\begin{eqnarray}
H_{ab} \equiv \int d Z  ({\overline F} F)_{ab}
\label{int1}
\end{eqnarray}
and the superfield  
\begin{eqnarray}
H_{0} \equiv \int d z ~tr (F{\overline F} )
\label{int2}
\end{eqnarray}
are integrals of the flows \p{gnls} as well. Furthermore,
at $n=0$ or $m=0$ the superfield  
\begin{eqnarray}
{\overline H}_0 \equiv \int d z ~tr ({\overline F} F)
\label{int11}
\end{eqnarray}
is also the integral of the flows \p{gnls}. 
The integrals $H_{ab}$ \p{int1} 
are fermionic (bosonic) ones when the indices $a,b$ belong to the
following ranges: $1\leq a \leq n$ and $n+1\leq b \leq n+m$, or
$n+1\leq a \leq n+m$ and $1\leq b \leq n$ ($1\leq a,b \leq n$, or 
$n+1\leq a,b \leq n+m$). Therefore, it is natural to suppose that besides
the bosonic flows (\ref{n1}--\ref{gnls}) and two fermionic flows of the
$N=2$ supersymmetry, originated from the superfield Hamiltonian 
$H_0$ \p{int2}, the $N=2$ unconstrained $(k|n,m)$--MGNLS
hierarchy possesses additional series of fermionic and bosonic flows,
related to the local fermionic and bosonic integrals $H_{ab}$ \p{int1} 
via the corresponding Hamiltonian structure, and that just
the algebra of these integrals is the $gl(n|m)$
superalgebra (see the discussion at the paragraph after eq. \p{help}).

\subsection{\bf Involutions}

We restrict our considerations to the case
when $iz$, $\theta$ and ${\overline \theta}$ are coordinates of the real $N=2$
superspace which satisfy the following standard complex conjugation
properties: 
\begin{eqnarray}
(iz,{\theta}, {\overline \theta})^{*}=(iz,{\overline \theta},{\theta})
\label{conjdef}
\end{eqnarray}
where $i$ is the imaginary unity. We will also use the standard
convention regarding complex conjugation $^*$ of products involving odd
operators and functions. In particular, if $\hbox{\bbd O}$ is some even
differential operator acting on a superfield $F$, we define the complex
conjugate of $\hbox{\bbd O}$ by $(\hbox{\bbd O}F)^*=\hbox{\bbd O}^*F^*$.
Then, in the case under consideration one can derive, for example, the
following relations
\begin{eqnarray}
&& (F_{Aa}{\overline F}_{bB})^{*}={\overline F}_{bB}^{*} F_{Aa}^{*},
\quad ({\epsilon},{\overline {\epsilon}})^{*}=
({\overline {\epsilon}},{\epsilon}), \quad
({\epsilon}{\overline {\epsilon}})^{*}={\epsilon}{\overline {\epsilon}},\quad
\nonumber\\
&&{\partial}^*=-{\partial}, \quad
({\epsilon}D, {\overline {\epsilon}}{\overline D})^{*}=
({\overline {\epsilon}}{\overline D},{\epsilon}D), \quad
(D{\overline D})^{*} = - {\overline D}D
\label{conjdef1}
\end{eqnarray}
which we use in what follows. Here, ${\epsilon}$ and ${\overline {\epsilon}}$
are constant odd parameters.

Direct verification shows that the evolution equations \p{gnls} admit two
different complex conjugations
\begin{eqnarray}
(F^{*},{\overline F})^{*} = ({\overline F}^{T},F^{T}), \quad
t^{*}_p= -t_p, \quad
(iz,{\theta},{\overline {\theta}})^{*}=
(iz,{\overline {\theta}}, {\theta})
\label{conj}
\end{eqnarray}
and
\begin{eqnarray}
(F, ~{\overline F})^{\star}= 
(X F{\cal I},
~{\cal I}{\overline F}X^{-1}),\quad  
t_p^{\star}= (-1)^{p}t_p, \quad
(iz,{\theta}, {\overline {\theta}})^{\star}=
(iz,{\overline {\theta}},{\theta})
\label{conj0}
\end{eqnarray}
where the matrices ${\cal I}$ and $X$ are defined in eqs. 
\p{matrixI} and \p{def}, respectively. Using eqs. \p{def} and
(\ref{conjdef1}--\ref{conj0}) one can derive the following involution 
properties of the matrix $X$:
\begin{eqnarray}
X^{*}=(X^{T})^{-1}, \quad X^{\star}=X^{-1}.
\label{propert}
\end{eqnarray}

\section{\bf Reductions and limited cases
of the $N=2$ unconstrained $(k|n,m)$--MGNLS hierarchy}

In this section we discuss different reductions and particular, limited
cases of the $N=2$ unconstrained $(k|n,m)$--MGNLS hierarchy as well as
their correspondence to known hierarchies.
  
\subsection{\bf Relations to the $N=2$ chiral $(k|n,m)$--MGNLS and $N=2$
multicomponent hierarchies}

For the case, when $F$ and ${\overline F}$ are constrained to be chiral
and antichiral rectangular matrix--valued $N=2$ superfields, i.e.
\begin{eqnarray}
D F=0, \quad {\overline D}~{\overline F} = 0,
\label{chiral}
\end{eqnarray}
respectively, the Lax operator $L$ \p{suplax} of the $N=2$ unconstrained
$(k|n,m)$--MGNLS hierarchy reproduces the Lax operator
of the $N=2$ chiral $(k|n,m)$--MGNLS hierarchy \cite{bks1} on the subspace
of the chiral wave function $\Psi$,
\begin{eqnarray}
L\Psi \equiv
\Bigl(I\partial - F {\overline F} -
F {\overline D}\partial^{-1} (D{\overline F})_0\Bigr)\Psi, \quad 
D\Psi =0.
\label{redchir}
\end{eqnarray}
Therefore, at the reduction \p{chiral} our hierarchy is equivalent to
the $N=2$ chiral $(k|n,m)$--MGNLS hierarchy.

In the very particular, scalar case, i.e. at $k=1$, the $N=2$
unconstrained $(1|n,m)$--MGNLS hierarchy is equivalent to the 
$N=2$ supersymmetric multicomponent hierarchy \cite{pop}. 
Indeed, in the new superfield basis $\{ {\widetilde F}, 
{\widetilde {\overline F}}\}$, defined as   
\begin{eqnarray} 
{\widetilde F}\equiv -\frac{i}{2}
Fe^{-{\frac{1}{4}{\partial}^{-1}(F {\overline F})}}, \quad 
{\widetilde {\overline F}}\equiv \frac{1}{2}
{\overline F}e^{+{\frac{1}{4}{\partial}^{-1}(F {\overline F})}}, 
\label{basispop}
\end{eqnarray}
the gauge--transformed Lax operator ${\widetilde L}$ \p{gauge}
\begin{eqnarray}
{\widetilde L}\equiv e^{-{\frac{1}{4}{\partial}^{-1}(F {\overline F})}} L 
e^{+{\frac{1}{4}{\partial}^{-1}(F {\overline F})}} = 
\partial + i {\widetilde F}[D,{\overline D}]{\partial}^{-1}
{\widetilde {\overline F}}
\label{poplax}
\end{eqnarray}
reproduces the Lax operator of the $N=2$ supersymmetric multicomponent
hierarchy \cite{pop}. At this point we would like to especially underline
that as concerns to the general, matrix case $k>1$, the $N=2$
unconstrained $(k|n,m)$--MGNLS hierarchy to our best knowledge is
introduced here for the first time. 
 
As a byproduct of this consideration, we have also established the
correspondence between the $N=2$ GNLS ($N=2$ chiral $(1|n,m)$--MGNLS)
hierarchy of ref. \cite{bks} and the $N=2$ multicomponent 
hierarchy of ref. \cite{pop}: the former hierarchy is
related to the latter by the reduction constraints \p{chiral}
and the basis transformation \p{basispop} (see, also the corresponding 
discussion in ref. \cite{ar}).

\subsection{\bf Bosonic limit}

Now we would like to discuss the bosonic limit of the $N=2$ unconstrained
$(k|0,m)$--MGNLS hierarchy using its second flow equations 
${\textstyle{\partial\over\partial t_2}}$ \p{gnls}
with the pure fermionic matrices $F, {\overline F}$, and establish their
relationship with the bosonic matrix $gl(2k+m)/(gl(2k)\times gl(m))$--NLS
equations introduced in \cite{fk}.

To derive the bosonic limit of the $N=2$ supersymmetric unconstrained
$(k|0,m)$--MGNLS hierarchy, let us define the matrix components of the
fermionic superfield matrices as
\begin{eqnarray}
&&f\equiv 
\left(\begin{array}{cc}  D~ F|\\  {\overline D}~ F| \end{array}\right), \quad 
\quad {\overline f}\equiv  \left(\begin{array}{cc} -{\overline D}~
{\overline F}|,&  D~{\overline F}| \end{array}\right),  \nonumber\\ 
&& {\psi} \equiv \left(\begin{array}{cc} F| \\{\overline D} D~ F|
\end{array}\right), \quad  {\overline {\psi}} \equiv \left(\begin{array}{cc} 
{\overline F}|, & D{\overline D} ~ {\overline F}|
\end{array}\right)
\end{eqnarray}
where $|$ means the $({\theta}, {\bar\theta})\rightarrow 0$
limit. So, $\psi$ and ${\overline \psi}$ are
fermionic matrix components, while $f$ and ${\overline f}$ are
bosonic ones. To get the bosonic limit we have to put the fermionic
matrices ${\psi}$ and ${\overline {\psi}}$ equal to zero. This leaves us
with the following set of matrix equations 
\begin{eqnarray}
{\textstyle{\partial\over\partial t_2}} f=f~''- f{\overline f}f, \quad
{\textstyle{\partial\over \partial t_2}} {\overline f} =
-{\overline f}~''+{\overline f}f{\overline f}
\label{bosgnls1}
\end{eqnarray}
for the bosonic matrix components $f$ and ${\overline f}$. The 
derived equations \p{bosgnls1} reproduce the bosonic matrix NLS
equations which can be elaborated via the $gl(2k+m)/(gl(2k)\times
gl(m))$--coset construction \cite{fk}. They can be viewed as the second
flow of the bosonic matrix NLS hierarchies with the matrix Lax operators
${\cal L}_1$
\begin{eqnarray}
{\cal L}_1= I\partial - \frac{1}{2}f \partial^{-1} {\overline f}
\label{boslax2}
\end{eqnarray}
which can easily be derived from the Lax operator \p{suplax} in the
bosonic limit. 

Thus we are led to the conclusion that the $N=2$ supersymmetric
unconstrained $(k|0,m)$--MGNLS hierarchy is the $N=2$ superextension 
of the bosonic matrix $gl(2k+m)/(gl(2k)\times gl(m))$--NLS
hierarchy. At this point let us remark the difference of the $N=2$
unconstrained $(k|0,m)$--MGNLS hierarchy comparing to the $N=2$
chiral $(r|0,m)$--MGNLS hierarchy: the latter corresponds to the $N=2$ 
superextension of the $gl(r+m)/(gl(r)\times gl(m))$--NLS hierarchy
\cite{bks1}. Therefore, at even value of $r$, $r=2k$, the $N=2$ chiral
$(2k|0,m)$--MGNLS hierarchy and $N=2$ unconstrained $(k|0,m)$--MGNLS
hierarchy are two different $N=2$ superextensions of the same bosonic
matrix hierarchy --- $gl(2k+m)/(gl(2k)\times gl(m))$--NLS hierarchy. It
seems these two $N=2$ superextensions are not equivalent in general
because of different length dimensions of their fermionic superfield
components, but, this question requires a more careful analysis which is
out of the scope of the present Letter.

\subsection{\bf Reduction of the $N=2$ unconstrained $(1|1,0)$--MGNLS
hierarchy} 

The $N=2$ unconstrained $(1|1,0)$--MGNLS hierarchy involves two bosonic
unconstrained $N=2$ superfields $F(Z)$ and ${\overline F}(Z)$, and
does not admit fermionic integrals \p{int1} (see the paragraph after eqs.
\p{int2}). Nevertheless, there is some hidden possibility for 
generating other fermionic integrals at its reduction which we discuss in
this subsection.

With this aim let us introduce the new superfield basis 
$\{ J,{\overline J} \}$, defined as\footnote{We assume the following
inverse length dimensions of the involved superfields: 
$[J]=[{\overline J}]=[{\overline F}]=1$ and $[F]=0$.}     
\begin{eqnarray} 
&&J\equiv \frac{1}{2} F {\overline F}-({\ln F})~', \quad
{\overline J}\equiv ({\ln F})~', \nonumber\\
&& {\overline F}= 2(J+{\overline J})e^{-{\partial}^{-1}{\overline J}},
\quad F= e^{{\partial}^{-1}{\overline J}}, 
\label{basiskdv}
\end{eqnarray}
in which the second and third bosonic flows \p{gnls} 
as well as the Hamiltonians \p{i1} and \p{int2} become
\begin{eqnarray}
&&{\textstyle{\partial\over\partial t_2}}{J} =
(-[D,{\overline D}~] J - J^2 + 2{\overline D}D {\overline J})~', \quad
{\textstyle{\partial\over\partial t_2}}{\overline J} =
(+[D,{\overline D}~] {\overline J} + {\overline J}^2 + 
2D{\overline D} J)~' 
\label{gnlskdv2}
\end{eqnarray}
and
\begin{eqnarray}
&&{\textstyle{\partial\over\partial t_3}}{J} =
[J~'' + 3 J[D,{\overline D}~] J + J^3 + 3 (DJ)({\overline D}J)-
3(D{\overline J}){\overline D}~{\overline J} +
3({\overline J}-J){\overline D}D{\overline J}]~',\nonumber\\ 
&&{\textstyle{\partial\over\partial t_3}}{\overline J} =
[{\overline J}~'' + 3 {\overline J}[D,{\overline D}~] {\overline J} + 
{\overline J}^3 - 3(DJ)({\overline D}J) + 
3(D{\overline J}){\overline D}~{\overline J} +
3({\overline J}-J)D{\overline D}J]~', 
\label{gnlskdv3}
\end{eqnarray}
as well as
\begin{eqnarray}
&&\quad \quad \quad \quad \quad \quad \quad 
H_{0}  = 2 \int dz ~J, \quad
H_2 = -2\int dZ ~(J^2-{\overline J}^2),
\nonumber\\&&H_3 = 6\int dZ~[J{\overline J}~'+
\frac{1}{2}(J+{\overline J})[D,{\overline D}](J+{\overline J})+
\frac{1}{3}(J+{\overline J})(J^2-J{\overline J}+{\overline J}^2)],
\label{ii1}
\end{eqnarray}
respectively, and admit the following complex conjugations
\begin{eqnarray}
(J,~{\overline J})^{*}=(J+(\ln (J+{\overline J}))~', 
~{\overline J}-(\ln (J+{\overline J}))~'), \quad
t^{*}_p= -t_p, \quad
(iz,{\theta},{\overline {\theta}})^{*}=
(iz,{\overline {\theta}}, {\theta}),
\label{conj1}
\end{eqnarray}
\begin{eqnarray}
(J, ~{\overline J})^{\star}= ({\overline J}, ~J), \quad 
t_p^{\star}= (-1)^{p}t_p, \quad
(iz,{\theta}, {\overline {\theta}})^{\star}=
(iz,{\overline {\theta}},{\theta})
\label{conjstar}
\end{eqnarray}
where at deriving eqs. (\ref{conj1}--\ref{conjstar}) we have used eqs.
(\ref{conj}--\ref{conj0}) and \p{basiskdv} as well as  
\begin{eqnarray} 
X= e^{-{\frac{1}{2}{\partial}^{-1}(F {\overline F})}}
\label{defX}
\end{eqnarray}
which results from eqs. \p{rese}, \p{solG} and \p{def} for the hierarchy
under consideration. 

The odd bosonic flows ${\textstyle{\partial\over\partial t_{2p+1}}}$ of
the hierarchy under consideration and the involution \p{conjstar} 
are consistent with the following reduction constraint: 
\begin{eqnarray}
J={\overline J}
\label{reduction}
\end{eqnarray}
which has the following form: 
\begin{eqnarray}
{\overline F} = -\Bigl(\frac{4}{F}\Bigr)~'
\label{reduction1}
\end{eqnarray}
in terms of the original superfields $F$ and ${\overline F}$ \p{basiskdv}.
Let us remark that in another superfield basis this reduction was
discussed in \cite{pop}, and it is equivalent to the requirement that even
Hamiltonians $H_{2p}$ \p{res} of the hierarchy are subjected equal to zero
(see, e.g. the Hamiltonian $H_2$ \p{ii1}). 

Now, one can easily verify that the composite fermionic superfield
\begin{eqnarray}
{\cal I}_{\frac{3}{2}} \equiv FD F~'- F~'D F 
\label{defmain}
\end{eqnarray}
satisfies the following important relation 
\begin{eqnarray}
{\textstyle{\partial\over\partial t_3}}{\cal I}_{\frac{3}{2}} =
\Bigl[F^2\Bigl(DJ~'' - 6 (DJ){\overline D}D J + 
JDJ~' - J~'DJ + J^2DJ\Bigr)\Bigr]~'
\label{diver}
\end{eqnarray}
on the constraint shell \p{reduction1}, therefore, both the third flow
${\textstyle{\partial\over\partial t_3}}$ \p{gnls} and the whole
reduced hierarchy possess the fermionic superfield integral  
\begin{eqnarray}
I_{\frac{1}{2}}= \int dz {\cal I}_{\frac{3}{2}} 
\label{reductionint}
\end{eqnarray}
as well as its complex conjugated quantity 
\begin{eqnarray}
I^{\star}_{\frac{1}{2}} =-\int dz 
{\cal I}^{\star}_{\frac{3}{2}},\quad 
{\cal I}^{\star}_{\frac{3}{2}} \equiv \frac{1}{F^4}
(F{\overline D}F~'-F~'~{\overline D}F),
\label{reductionintstar}
\end{eqnarray}
\begin{eqnarray}
{\textstyle{\partial\over\partial t_3}}{\cal I}^{\star}_{\frac{3}{2}} =
\Bigl[ \frac{1}{F^2}\Bigl({\overline D}J~'' + 
6 ({\overline D}J)D{\overline D} J -  J{\overline D}J~' + 
J~'~{\overline D}J + J^2{\overline D}J\Bigr)\Bigr]~'
\label{diverstar}
\end{eqnarray}
which are nonlocal and even nonpolynomial in the basis 
$\{ J,{\overline J} \}$ \p{basiskdv} where their densities are
\begin{eqnarray}
{\cal I}_{\frac{3}{2}}= e^{+2{\partial}^{-1}J}DJ, \quad
{\cal I}^{\star}_{\frac{3}{2}} = e^{-2{\partial}^{-1}J}
{\overline D}J.  
\label{reductionintkdv}
\end{eqnarray}
Here, the subscripts denote inverse length dimensions,
and when calculating eqs. (\ref{reductionintstar}--\ref{diverstar}) we
have applied the complex conjugation \p{conj0} with the function $X$
\p{defX} (restricted to the constraint shell \p{reduction1}) to eqs.
(\ref{defmain}--\ref{reductionint}),  
\begin{eqnarray}
F^{\star}= \frac{1}{F}, \quad J^{\star} = J, \quad 
t_p^{\star}= (-1)^{p}t_p, \quad
(iz,{\theta}, {\overline {\theta}})^{\star}=
(iz,{\overline {\theta}},{\theta}).
\label{conjstarF}
\end{eqnarray}
We would like to remark that in another field basis the superfield
components of the integrals \p{reductionintkdv} were derived recently in
\cite{ker,ker1} by a tedious symmetry analysis using computer
calculations, but their $N=2$ superfield structure and origin were not
clarified there. 

On the constraint shell \p{reduction} the third flow \p{gnlskdv3}
and the third Hamiltonian $H_3$ \p{ii1} become
\begin{eqnarray}
{\textstyle{\partial\over\partial t_3}}{J} =
[J~'' + 3 J[D,{\overline D}~] J + J^3]~'
\label{gnlskdvn2}
\end{eqnarray}
and
\begin{eqnarray}
H_3 = 12\int dZ~(J[D,{\overline D}]J+\frac{1}{3}J^3),
\label{iii1}
\end{eqnarray}
respectively, and one can easily recognize that they reproduce the third
flow and the third Hamiltonian of the $N=2$ supersymmetric ${\alpha}=1$
KdV hierarchy possessing the $N=2$ superconformal algebra as the second
Hamiltonian structure \cite{lm}.
Therefore, we are led to the conclusion that the $N=2$ ${\alpha}=1$ KdV
hierarchy possesses hidden fermionic superfield
integrals \p{reductionintkdv} as well as corresponding
flows. Actually, the integrals (\ref{reductionint}--\ref{reductionintstar})
are only the first representatives of the series of nonlocal fermionic and
bosonic superfield integrals arising at the reduction \p{reduction1}.
Their more detailed consideration as well as  
their role for more deep understanding and more detail description of the
$N=2$ ${\alpha}=1$ KdV hierarchy is out of the scope of the present Letter
and will be discussed in \cite{ks}.  

To close this section we would like to remark that the established
relation between the $N=2$ unconstrained $(1|1,0)$--MGNLS and $N=2$
${\alpha}=1$ KdV hierarchies allows to derive the following nice formula
for the flows ${\textstyle{\partial\over\partial t_{2p+1}}}$ 
and the corresponding Hamiltonian densities ${\cal H}_{2p+1}$ of the
latter hierarchy
\begin{eqnarray}
&&{\textstyle{\partial\over\partial t_{2p+1}}}J = {\cal H}^{~'}_{2p+1} =
(res(L^{2p+1}))~'\equiv (res({\widetilde L}^{2p+1}))~', \nonumber\\
&& {\widetilde L}\equiv e^{-({\partial}^{-1}J)} Le^{({\partial}^{-1}J)}
=\partial + [D,{\overline D}]{\partial}^{-1}J
\label{suplaxconstflow}
\end{eqnarray}
where when deriving we have used the equations \p{suplaxconst}, \p{res},
\p{basiskdv}, (\ref{reduction}--\ref{reduction1}) as well as the identity   
\begin{eqnarray}
res(f^{-1}\hbox{\bbd O}f) \equiv res(\hbox{\bbd O})
\label{identity}
\end{eqnarray}
which is valid for any $N=2$ pseudo--differential operator $\hbox{\bbd O}$ and
superfield $f$.

\section{\bf Summary} 

In this Letter we have proposed a wide class of new $N=2$ supersymmetric 
hierarchies --- the $N=2$ supersymmetric unconstrained $(k|n,m)$--MGNLS
hierarchy --- by exhibiting the corresponding super 
Lax--pair representation \p{suplax} in terms of matrix--valued $N=2$
unconstrained superfields. Then we have explicitly calculated its 
first nontrivial flows (\ref{n1}--\ref{gnls}) and conserved
quantities (\ref{res}--\ref{int11}). Furthermore we have constructed its
two different admissible involutions (\ref{conj}--\ref{conj0}).
Then we have discussed its different limited cases and reductions as well
as the correspondence to already known hierarchies. Finally we have
analysed its simplest representative -- the $N=2$ unconstrained
$(1|1,0)$--GNLS hierarchy -- and the reduction to the $N=2$ supersymmetric
${\alpha}=1$ KdV hierarchy, and a rich symmetry structure of the latter is
uncovered. 

{}~

\noindent{\bf Acknowledgments.}
A.S. is grateful to H. Aratyn for useful discussions at the early stage
of this study and to University of Twente for the hospitality extended to
him during this research. This work was partially supported by the grants
NWO NB 61-491, FOM MF 00/39, RFBR 99-02-18417, RFBR-CNRS 98-02-22034, PICS
Project No. 593, Nato Grant No. PST.CLG 974874 and the Heisenberg-Landau
program.

\end{document}